\documentclass[12pt]{article}
\usepackage{amssymb}

\def\beq#1{\begin{equation}\label{#1}}
\def\eeq{\end{equation}}
\def\beqa#1{\begin{eqnarray}\label{#1}}
\def\eeqa{\end{eqnarray}}

\def\comment#1{\relax}

\begin{document}
\begin{center}
{\bf\large Gamma-ray Bursts and Hypernovae}\\
{\large K.A.Postnov}\\
{\it Sternberg Astronomical Institute, Moscow, Russia, 
and University of Oulu, Finland}
\end{center}
\vskip\baselineskip

{\it Summary.} In these proceedings, I discuss recent 
progress in understanding the nature of cosmic gamma-ray bursts 
(GRB), with the focus on the apparent relation of several GRBs with 
an energetic subclass of stellar explosions, type Ib/c core-collapse
supernovae. This relation provides the strong case that the GRB phenomenon 
is connected with the final stages of massive star evolution 
and possibly with the formation of neutron stars and black holes.
I speculate that intrinsically faint, apparently spherically symmetric 
nearby GRB 980425 and 031203 associated
with bright hypernovae SN 1998bw and SN 2003lw, respectively, can signal
the formation of a neutron star  in the end of gravitational collapse,
while the bulk of cosmological GRBs with a universal
energy release of $\sim 10^{51}$ ergs in narrow-collimated jets 
are produced when a black hole is formed. In the former case, the 
energy source of GRB is the neutron star rotational energy; in the latter 
case the GRB energy is due to non-stationary accretion onto the black hole.      

\section{Introduction}
GRBs have remained in the focus of modern astrophysical 
studies for more than 30 years. After the discovery of GRB afterglows
in 1998 (Costa et al. 1998), the model of GRB as being due to 
a strong explosion with isotropic energy release of $10^{53}$ 
ergs in the
interstellar medium (originally proposed by Rees and Meszaros (1992))
became widely recognized. Various aspects of GRB phenomenology 
are discussed in many reviews: observational and theoretical 
studies are summarized in Hurley et al. (2003), 
first observations of afterglows are
specially reviewed in van Paradijs et al. (2000), GRB theory 
is extensively discussed in Meszaros (2002). 

A widely used paradigm for GRBs is the so-called fireball model (e.g. Piran
2004 and references therein). In this model, the energy 
is released in the form of thermal energy
(its initial form is usually not specified) near the compact 
central source (at distances and is mostly converted into leptons and photons
(the fireball itself). The outflow is formed driven by the high 
photon-lepton pressure (generically in the form of two
oppositely directed narrow collimated jets).
The fireball internal energy is converted to the 
bulk motion of ions so that relativistic speed with high Lorentz-factors 
(typically, $\Gamma > 100$) is achieved during the initial stage of 
the expansion; the ultrarelativistic motion is in fact dictated by the need to 
solve the fireball 
compactness problem (see Blinnikov 2000 for a detailed discussion
and references).
The kinetic motion of ions is reconverted back 
into heat in strong collisionless relativistic shocks at typical 
distances of $10^{12}$ cm. Assuming 
the appropriate turbulence 
magnetic field generation and particle acceleration
in the shocks, energy thermalized in the shocks is emitted via 
synchrotron radiation of accelerated electrons 
(see Waxman 2003), which is identified
with the GRB emission. A shell of ultrarelativistically  moving 
cold protons produce a blast wave in the surrounding medium, forming an 
external shock propagating outward and reverse shock propagating inward the
explosion debris. Most energy of explosion is now carried by the external
shock which decelerates in the surrounding medium. Assuming magnetic field
generation and particle acceleration in the external shock, the afterglow 
synchrotron emission of GRB is produced. 
Note that at this stage the memory of the initial
explosion conditions is cleaned, and the dynamical evolution of the 
external shock is well described by the Blandford-McKee (1975) 
self-similar solution, a relativistic analog of the Sedov-von-Neumann-Taylor 
solution for strong point-like explosion. 
This explains the success in modeling the GRB afterglow
spectral and temporal behavior in the framework of  
the synchrotron model (Wijers et al. 1997), irrespective of 
the actual nature of the GRB explosion. 

Indeed, there is no consensus thus far about the origin of
the GRB emission itself. The fireball model meets some important 
problems (for example, baryon contamination of the fireball, the
microphysics of magnetic field generation and particle acceleration 
in collisionless ultrarelativistic shocks etc., see a more detailed list 
in Lyutikov and Blandford (2003)). In the last paper an 
alternative to the fireball model was proposed in which large-scale 
magnetic fields are dynamically important. Whether the GRB jets are hot
(fireball model) or cold (electromagnetic model) remains to be determined
from future observations. Here crucial may be spotting the very early 
GRB afterglows and measuring polarization of prompt GRB emission (see
Lyutikov (2004) for the short-list of the electromagnetic model predictions).

Here we focus on the observed association of 
GRBs with an energetic subclass of core-collapse supernovae (SNe), 
type Ibc SNe, which with each new finding provides an increasing
evidence that the GRB phenomenon is related to the evolution of
most massive stars and formation of stellar-mass black holes (BH). 

\section{Supernova - GRB connection} 

\subsection{Theoretical grounds: the collapsar model}

The connection of GRBs with stellar explosions was first proposed
theoretically. Woosley (1993) considered a model 
of accretion onto a newly formed rotating black hole to power
the GRB fireball. The progenitor to GRB in this model is a rapidly rotating
Wolf-Rayet (WR) star deprived of its hydrogen and even helium envelop 
due to powerful stellar wind or mass transfer in a binary system. 
Dubbed by Woosley himself as "failed type Ib supernovae", this model is now 
called the collapsar model (MacFadyen and Woosley, 1999). 
In this model, a  massive ($\gtrsim 25 M_\odot$) rotating star 
with a helium core $\gtrsim 10 M_\odot$  
collapses to form a rapidly rotating BH with 
mass $\gtrsim 2-3 M_\odot$. The accretion disk from the 
presupernova debris around the BH is assumed to be the energy 
source for GRB and is shown to be capable of providing the prerequisite 
$10^{51}-10^{52}$ ergs via viscous dissipation into neutrino-antineutrino
fireball. The energy released is
assumed to be canalized in two thin antiparallel jets penetrating
the stellar envelop. Another possible energy source in the collapsar 
model could be the electromagnetic (Poynting-dominated) beamed 
outflow created via MHD processes, much alike what happens 
in the active galactic nuclei powered by accretion onto a supermassive
BH. The estimates show that  
the Blandford-Znajek (BZ) (1977) process in the collapsar model
(e.g. Lee et al. 2000) can be a viable candidate for 
the central engine mechanism for 
GRBs, provided somewhat extreme values for BH spin (the Kerr parameter
$a\sim 1$) and magnetic field strength in the inner accretion 
disk around the BH ($B\sim 10^{14}-10^{15}$ G). In that case the 
rotating energy of BH (up to $0.29 M_{bh}c^2$ for $a=1$) 
is transformed to 
the Poynting-dominated jet with energy sufficient to 
subsequently produce GRB.  

Another source of energy in the collapsar model could be the rotation energy
of a rapidly spinning neutron star with high magnetic field (magnetar), as
originally proposed by Usov (1992). As in the BZ-based models, the GRB jets
are Poynting-dominated. Lyutikov and Blandford (2003) develop the
electromagnetic model, which postulates that the rotating energy of the GRB
central engine is transformed into the electromagnetic energy (for example,
in a way similar to the Goldreich-Julian pulsar model) and is stored in a
thin electromagnetically-dominated "bubble" inside the star. The bubble
expands most rapidly along the rotational axis, breaks out of the stellar
envelopes and drives the ultrarelativistic shock in the circumstellar
material. In contrast to the synchrotron GRB model, here GRB is produced
directly by the magnetic field dissipation due to current-driven
instabilities in this shell after the breakout. The energy transfer to GRB
is mediated all the way by electromagnetic field and not by the ion bulk
kinetic energy. As we noted in the Introduction, it remains to be checked by
observations whether the EM or fireball model for GRB emission is correct.

\subsection{Observational evidence: GRB-supernova associations} 

First hint on the association of GRBs with SNe came from 
the apparent time coincidence (to within about a day) 
of GRB 980425 with a  
peculiar supernova SN 1998bw (Galama et al. 1998).
SN 1998bw occurred in a spiral arm of  
nearby (redshift $z=0.0085$, distance $\sim 40$ Mpc) spiral galaxy ESO 184-G82. 
Such a close location of GRB 980425 rendered it a 
significant outliers by (isotropic) energy release 
$\Delta E_{iso}\approx 10^{48}$ erg from the bulk of 
other GRBs with known energy release, and even from 
a beaming-corrected mean value of GRB energies of $\sim 10^{51}$ erg
(Frail et al. 2001).  

Now the most convincing evidence for GRB-SN association 
is provided by spectroscopic observations of late 
GRB afterglows. Among them is a bright GRB 030329 associated with SN 2003dh
(Hjorth et al. 2003, Stanek et al. 2003, Matheson et al. 2003, Mazzali et al. 2003, 
Kawabata et al. 2003). Spectral observations of the optical
afterglow of this GRB revealed the presence of
thermal excess above non-thermal power-law continuum  typical
for GRB afterglows. Broad absorption troughs which became more
and more pronounced as the afterglow faded indicated the presence of
high-velocity ejecta similar to those found in 
spectra of SN 1998bw. Despite these strong evidences, 
there are some facts which cannot be explained by simple
combination of the typical SN Ibc spectrum and non-thermal power-law continuum.
For example, the earliest spectroscopic observations of GRB 030329 of 
optical spectra taken on the 6-m telescope SAO RAS 
10-12 hours after the burst (Sokolov et al. 2004) showed the presence
of broad spectral features which could not be produced by a SN 
at such an early stage. The complicated shape of the 
optical light curve of this GRB with many rebrightenings 
(Lipkin et al. 2004) and 
polarization observations made by VLT (Greiner et al. 2003) suggest 
a clumpy circumburst medium and require additional refreshening of
shocks (if one applies the synchrotron model, e.g. Granot et al. (2003)).   

Another recent example of GRB-SN connection is provided by another nearby GRB
031203. This GRB is one of the closest ($z=0.105$) known GRBs and is found
to be intrinsically faint, $\Delta E_{iso}\sim 10^{50}$ ergs (Watson et al. 
2004, Sazonov et al. 2004)\footnote{A bright soft X-ray flux was inferred from XMM observations
of evolving X-ray halo for this burst (Vaughan et al. 2004), making it 
an X-ray rich GRB (Watson et el. 2004); this point of view was argued by
Sazonov et al. (2004).}. 
The low energy release in gamma-rays is confirmed by the 
afterglow calorimetry derived
from the follow-up radio observations (Soderberg et al. 2004) and
allows this GRB to be considered as an analog to GRB 980425. It is important 
that the low energy release in these bursts can not be ascribed to the 
off-axis observations of a "standard" GRB jet (unless one assumes 
a special broken power-law shape of GRB luminosity function, 
see Guetta et al. 2004). However, a
bright type Ib/c supernova SN 2003lw was associated with GRB 031203 as
suggested by the rebrightening of the R light curve peaking 18 days after
the burst and broad features in the optical spectra taken close to the
maximum of the rebrightening (Cobb et al 2004, Thomsen et al. 2004, 
Malesani et al. 2004, Gal-Yam et al. 2004).

The comparison of radio properties of 33 SNe type Ib/c with those 
of measured radio GRB afterglows allowed Berger et al. (2003) 
to conclude that not more than few per cents of SNe type Ib/c could be 
associated with GRBs, which explains the observed small 
galactic rate of GRBs. However, it still remains to be studied how much 
intrinsically faint GRBs like 980425 and 031203 can contribute to the
total GRB rate.

\section {Hypernovae}

Core-collapse supernovae with kinetic energy of the ejecta 
$\sim 10-30$ times as high as the standard 1 foe ($1\hbox{foe}=10^{51}$ erg)
are now collectively called "hypernovae". The term was introduced by 
B. Paczynski shortly after the discovery of first GRB afterglows in 1997 
by the Beppo-SAX satellite (Paczynski, 1998) based on qualitative 
analysis of possible evolutionary ways leading to cosmic GRB explosions. 

SN 1998bw was exceptionally 
bright compared to other Ib/c SNe 
(the peak bolometric luminosity of order $10^{43}$ erg/s, comparable to
the SN Ia peak luminosities). This points to the
presence of a substantial amount of $^{56}$Ni isotope, the radioactive decay
thereof being thought to power the early SN light curves.
The spectra and light curve of 
SN 1998bw was modeled by the explosion of a bare C+O of a very massive
star that has lost its hydrogen and helium envelopes with a kinetic energy 
more than ten times typical SNe energies (Iwamoto et al. 1998), 
and they called SN 1998bw  
a hypernova.

Since then several 
other SNe were classified as SN 1998bw-like 
hypernovae by their spectral
features and light curves: SN 1997ef, SN 2002ap, SN 2003dh/GRB030329, SN 2003lw/031203. Recently, SN 1997dq was dubbed a hypernova 
by its similarity with SN 1997ef (Mazzali et al. 2004).   
   
Extensive numerical modeling of light curves and spectra
of hypernovae (see Nomoto et al. 2004 for a recent review) 
confirmed the need of atypically high for core-collapse SNe 
mass of nickel ($\sim 0.1-0.5
M_\odot$) to be present in the ejecta in order to
explain the observed hypernova properties. 
The rapid rise in of the observed light curves of the "canonical" SN 1998bw 
requires a substantial amount of $^{56}$Ni to be present near the surface.
This strongly indicates the important role of 
mixing during the explosion as
nickel is synthesized in deep layers during a spherical explosion.  
This fact can serve as an additional evidence for non-spherical
type Ic explosions. Generally, the asphericity appears to be a ubiquitous feature of core-collapse supernovae. For example, spectropolarimetry 
of SN spectra (Leonard and Filippenko 2004) indicates the increasing
polarization degree for type Ib/c SNe compared to 
classical type II core-collapse SNe with rich hydrogen envelope (SN IIp), 
in which asymmetry appears to be dumped by the addition of envelope material.       

Spectral modeling suggests (Nomoto et al. 2004) 
that the broad-band spectral features generally seen in early and maximum light
of hypernovae signal very rapid photospheric expansion. For example, Nomoto
notes the very unusual for other SNe fact 
that OI ($\lambda =7774 A$) and CaII IR (at $\lambda \sim 8000 A$)
absorption
lines merge into a single broad absorption in early spectra of SN 1998bw,
which indicates a very large velocity of the ejecta (the line separation $\sim
30000$ km/s).  

In general, varying (a) the progenitor C+O core mass from 2 to $\sim 14$ solar
masses, choosing (2) the appropriate mass cut (corresponding to the mass of the
compact remnant, a neutron star or black hole $M_c=1.2-4 M_\odot$), 
and (3) mass of $^{56}$Ni isotope ($\sim 0.1-0.5 M_\odot$) and its 
mixing allow Nomoto et al. (2004) to obtain the observed spectra and light curves of hypernovae.

This analysis suggests a possible
classification scheme of supernova explosions. In this scheme, core collapse
in stars with initial main sequence masses 
$M_{ms}<25-30 M_\odot$ leads to the formation of neutron stars, while more
massive stars end up with the formation of black holes. Whether or not the
collapse of such massive stars is associated with powerful hypernovae
("Hypernova branch") or faint supernovae ("Faint SN branch") can depend on
additional ("hidden") physical parameters, such as the presupernova 
rotation, magnetic fields.
(Ergma and van den Heuvel 1998), or the GRB progenitor being a 
massive binary system component (Tututkov and Cherepashchuk 2003). The
need for other parameters determining the outcome of the core collapse also
follows from the observed continuous distribution of C+O cores of massive
stars before the collapse and strong discontinuity between masses of
compact remnants (the mass gap between neutron stars and black holes)
(Cherepashchuk 2001). The mass of $^{56}$Ni synthesized in core
collapse also appears to correlate with $M_{ms}$. In ordinary SNe (like
1987a, 1993j, 1994i), $M_{Ni}=0.08\pm 0.03 M_\odot$, but for hypernovae this
mass increases up to $\sim 0.5 M_\odot$ for the most energetic events.

Another important consequence of hypernovae can be different 
explosive nucleosynthesis products. Here the most pronounced 
features are larger 
abundances (relative to the solar one) 
of Zn, Co, V and smaller abundance of Mn, Cr, the enhanced ratios of
$\alpha$-elements, and large ratio of Si, S relative  to oxygen
(see Nomoto et al. 2003 for more detail). 

\section{Progenitors of GRBs}

The GRB-SN connection leads to the almost generally accepted 
concept that massive stars that lost their envelopes are progenitors of
long GRBs (this limitation is due to the fact that predominantly long GRBs
with duration > 2 s can be well localized on the sky and provide
rapid alerts for follow-up multiwavelength observations). For short 
single-pulsed GRBs (a quarter of all bursts, see e.g. catalog by Stern et al. 
2001) the binary NS+NS/NS+BH merging hypothesis 
(Blinnikov et al. 1984, Ruffert and Janka 1999, Janka et al. 1999) remains viable. 

As we already noted, the emerging evidence is that there are intrinsically
faint, single-pulsed, apparently spherically-symmetric GRBs (980425, 031203)
associated with strong hypernovae. These hypernovae require 
maximal amount of nickel to be synthesized in explosion 
and large kinetic energies. 
On the other hand, another unequivocal 
hypernova SN 2003dh, associated with the "classical" GRB 030329, can be
modeled with exceptionally high kinetic energy 
($4\times 10^{52}$ ergs) but smaller amount of nickel ($\sim 0.35 M_\odot$)
and smaller mass of the ejecta ($8-10 M_\odot$) (Mazzali et al. 2003). These
parameters were obtained assuming spherical symmetry, which is of course
not the case for GRB 030329. But if this tendency is real and will be confirmed
by later observations, we can return to our hypothesis 
(Postnov and Cherepashchuk 2001) that there 
should be distinct classes of GRBs according to what is the 
final outcome of collapse of the CO-core of a massive star. 
If the collapse ends up with the formation of a neutron star, 
intrinsically faint smooth GRB could be produced 
and heavy envelope is ejected in 
the associated SNIb/c explosion. The GRB energy in this case 
can be essentially the rotation energy of the neutron star $\sim 10^{49}-10^{50}$ ergs, 
as in the electromagnetic model by Usov (1992).  
If a BH is formed, a lighter envelope is ejected 
with accordingly smaller amount of nickel and possibly 
with higher kinetic energy, 
and more energetic, highly variable 
GRB with a "universal" jet structure (Postnov et al. 2001) appears
fed by non-stationary accretion onto the BH. 

The GRB energy dichotomy can be also interpreted in another, more exotic way 
requiring a new physics. For example, it was recently suggested 
(Gianfanga et al. 2004) that 
ultramassive axions in the mirror world with the Peccei-Quinn scale $f_a\sim 
10^4-10^6$ GeV and mass $m_a~1$ MeV can be produced in the gravitational collapse
or merging of two compact stars. The axions tap most of the released 
energy and can decay $\sim 1000$ km away mostly into visible 
electron-positron pairs (with $100\%$ conversion efficiency)
thus creating the initial GRB fireball. The estimates show that successful short GRBs 
can be obtained in compact binary coalescences, while long GRBs can 
be created in collapsars. In extended SNII progenitors, this energy may 
help the mantle ejection. In compact CO-progenitors for SN Ib/c axions decay inside
the star, so depending on the stellar radius weaker or stronger GRBs associated with 
SNe type Ib/c explosions can be observed. In this picture again the collapse with the 
formation of a neutron star or BH may have different signatures.

\section{Conclusions}

There are several unequivocal associations of cosmic GRBs with peculiar
very energetic type Ib/c SNe (hypernovae). The two closest GRBs 
discovered so far (989425 and 031203) proved to be intrinsically weak
compared to the bulk of other GRBs with measured redshifts. 
They both show a single-peak smooth 
gamma-ray light curve with no signs of jet-induced breaks 
in the afterglows. In the third (most strong) case of the GRB-SN association, 
GRB 030329/SN 2003dh, the GRB light curve is two-peaked, the afterglows
show evidence for jet. Modeling of the underlaid hypernovae light curve and spectra 
 revealed the first two cases to require smaller kinetic energies but higher 
mass of the ejecta and the amount of the synthesized nickel than SN 2003dh. We
propose that the  
tendency "weaker, more spherically symmetric GRB - stronger hypernova" is 
due to the formation of a NS in the case of 
weak GRBs and of a BH in the case of strong variable GRBs 
as the final outcome of the core collapse. In the NS case the GRB energy comes from
the rotational energy of neutron star and is possibly mediated by the 
electromagnetic field. When BH is formed the GRB energy source is 
the gravitational energy 
released during non-stationary accretion onto the black hole or the black hole 
rotation.  
 
We are sure that the increasing statistics of GRB/SNe in the nearest future
obtained with new GRB-dedicated space 
missions like SWIFT will tell us much more on the nature of GRBs and their 
progenitors.       

\bigskip

The author acknowledges the Organizing Committee of the QUARKS-2004 conference for
financial support. The work was also supported by RFBR grants 02-02-16500, 
03-02-17174 and 04-02-16720.


\begin{thebibliography}{50}

\bibitem{}
Berger E. et al. 2003, ApJ, 599, 408 

\bibitem{}
Blandford R., Znajek R.L., 1977, MNRAS 179, 433

\bibitem{}
Blinnikov S.I. et al., 1984, SvA Letters, 10, 177

\bibitem{}
Blinnikov S.I., 2000, Surveys High Energy Phys., 15, 37 (astro-ph/9911138)

\bibitem{}
Cherepashchuk A.M., 2001, Astron. Rep., 45, 120

\bibitem{}
Cobb B.E. et al., 2004, ApJ, 608, L93

\bibitem{}
Costa E. et al. 1997, Nature, 387, 783

\bibitem{}
Ergma E., van den Heuvel E.P.J., 1998, Astron. Astrophys., 331, L29

\bibitem{}
Frail D.A. et al., 2001, ApJ, 562, L55

\bibitem{}
Gal-Yam A. et al., 2004, ApJ, 609, L59

\bibitem{}
Gianfanga L. et al., 2004, preprint hep-ph/0409185

\bibitem{}
Granot J., Nakar E., Piran T., 2003, Nature, 426, 138

\bibitem{}
Greiner J. et al., 2003, Nature, 426, 257

\bibitem{}
Guetta D. et al., 2004, astro-ph/0409715

\bibitem{}
Iwamoto K. et al., 1998, Nature, 395, 672

\bibitem{}
Janka H.-T. et al., 1999, ApJ, 527, L39

\bibitem{}
Hjorth J. et al., 2003, Nature, 423, 847

\bibitem{}
Hurley K., Sari R., Djorgovki S.G., in Compact Stellar X-ray Sources,
Eds. W. Lewin and M. van der Klis, Cambridge Univ. Press, 2003
(astro-ph/0211620)

\bibitem{}
Leonard D.C., Filippenko A.V. 2004, in Supernovae as 
Cosmological Lighthouses, ed. M.Turrato et al., AIP Conf. ser.,
in press (astro-ph/0409518) 


\bibitem{}
Lipkin Y.M. et al. 2004, ApJ 606, 381

\bibitem{}
Lyutikov M., 2004, astro-ph/0409489

\bibitem{}
Lyutikov M., Blandford R., 2003, astro-ph/0312347

\bibitem{}
Lee H.K., Wijers R.A.M.J., Brown H.A., 2000, Phys. Rep. 325, 83

\bibitem{}
MacFadyen A., Woosley S., 1999, ApJ 524, 262

\bibitem{}
Malesani D. et al. 2004, ApJ, 609, L5

\bibitem{}
Mazzali P.A. et al. 2003, ApJ, 599, L95 

\bibitem{}
Mazzali P.A. et al. 2004, ApJ in press (astro-ph/0409575) 


\bibitem{}
Meszaros P., 2002, ARAA, 40, 137

\bibitem{}
Nomoto K. et al. 2003, Progr. Theor. Phys. Suppl., 151, 44

\bibitem{}
Nomoto K. et al. 2004, in Stellar Collapse, Ed. C.L.Fryer (Astrophysics
and Space Science, Kluwer) (astro-ph/0308136)


\bibitem{}
Paczynski B., 1998, ApJ, 494, L45

\bibitem{}
Piran T., 2000, Phys. Rep., 333, 529

\bibitem{}
Piran T., 2004, Rev. Mod. Phys., in press (astro-ph/0405503)

\bibitem{}
Postnov K.A., Prokhorov M.E., Lipunov V.M., 2001, Astron. Rep., 45, 236
(astro-ph/9908136) 

\bibitem{}
Postnov K.A., Cherepashchuk A.M., 2001, Astro. Rep., 45, 517

\bibitem{}
Rees M. J., Meszaros P. 1992, MNRAS, 258, 41P

\bibitem{}
Ruffert M., Janka H.-T., 1999, Astron. Astrophys., 344, 573

\bibitem{}
Sokolov V.V. et al. 2004, Bull. Special Astrophys. Obs. RAS, 56, 5

\bibitem{}
Stern B.E. et al., 2001, ApJ, 563, 80

\bibitem{}
Thomsen B. et al., 2004, Astron. Astrophys., 419, L21

\bibitem{}
Tutukov A.V., Cherepashchuk A.M., 2003, Astron. Rep., 47, 386

\bibitem{}
Usov V.V., 1992, Nature 357, 472

\bibitem{}
van Paradijs J., Kouveliotou C., Wijers R.A.M.J., 2000, ARAA, 38, 379

\bibitem{}
Vaughan S. et al., 2004, ApJ, 603, L5
 
\bibitem{}
Watson D. et al., 2004, ApJ, 605, L101

\bibitem{}
Wijers R.A.M.J., Rees M.J., Meszaros P., 1997, MNRAS 288, L51 

\end{thebibliography}
\end{document}